\documentclass[aps,preprint,preprintnumbers,amsmath,amssymb,nofootinbib]{revtex4}
\usepackage{amssymb}
\usepackage{lscape}
\usepackage{amsmath}
\usepackage{rotating}
\usepackage{bm}
\usepackage{graphicx}
\usepackage{epsfig}
\begin{document}
\title {Dyonic Reissner-Nordstr\"om black hole: extended Dirac quantization from  5D invariants.}
\author{$^{2}$
Jes\'us Mart\'{\i}n Romero\footnote{E-mail address:
jesusromero@conicet.gov.ar}, $^{1,2}$ Mauricio Bellini
\footnote{E-mail address: mbellini@mdp.edu.ar} }
\address{$^1$ Departamento de F\'isica, Facultad de Ciencias Exactas y
Naturales, Universidad Nacional de Mar del Plata, Funes 3350, C.P.
7600, Mar del Plata, Argentina.\\
$^2$ Instituto de Investigaciones F\'{\i}sicas de Mar del Plata (IFIMAR), \\
Consejo Nacional de Investigaciones Cient\'ificas y T\'ecnicas
(CONICET), Mar del Plata, Argentina.}
\begin{abstract}
 The aim of present work is to extend the application of Weitzeb\"{o}ck Induced Matter Theory (WIMT) to a dyonic Reissner-Nordstr\"{o}m Black Hole (RNBH), by proposing a condition compatible with a quantization relation between gravitational mass and both magnetic and electric charges from a geometric product defined as an invariant in 5D.
\end{abstract}
\maketitle

\section{Introduction}

In a previous work \cite*{1} we have developed an extension of the Induced Matter Theory (IMT) \cite*{wesson}
in presence of a Weitzenb\"{o}ck geometry, called Weitzenb\"{o}ck Induced Matter Theory (WIMT).
The spirit of IMT is that ordinary matter and physical fields present in our $4D$ universe can be geometrically induced from a $5D$ space-time which
is at least Ricci-flat (in the sense of Riemannian connections) with a non-compact extra dimension, so that the 5D physical vacuum defined by the Ricci-flatness condition supports the Campbell-Magaard embedding theorem as a particular case of $^{(5D)}R_{ab}=\lambda\, ^{(5D)}g_{ab}$, in which $\lambda=0$. This makes possible to define a 5D
apparent vacuum that determines the equations of motion for the fields of the theory. The aim of WIMT is to use an alternative description,
based in the Weitzenb\"{o}ck geometry, to apply an IMT-like formalism for any $5D$ space-time even if this is not flat in the Riemannian sense.
With the WIMT one can make the geometrical description of a problem from a non-flat (but torsion-less) Riemannian manifold using the Levi-Civita connections, into a Weitzenb\"{o}ck's geometrical description with a Riemann-Weitzenb\"{o}ck null tensor, because the Riemann-Weitzenb\"{o}ck tensor is the Riemann curvature tensor expressed in terms of Weitzenb\"{o}ck connections: $^{(W)}\Gamma^a_{\,\,\,dc}$. In a coordinate (holonomic) basis the Riemann-Weitzenb\"{o}ck curvature takes the form
\begin{eqnarray}
^{(W)}R^a_{bcd}&=&^{(W)}\Gamma^a_{dc\,,\,b} -^{(W)}\Gamma^a_{db\,,\,c} +
\,^{(W)}\Gamma^n_{dc} \,^{(W)}\Gamma^a_{nb}\nonumber \\
&-&\,^{(W)}\Gamma^n_{db}\,^{(W)}\Gamma^a_{nc}.
\end{eqnarray}
In this case we can use the IMT tools to induce $4D$ effective space-time dynamics by making a foliation over the Riemann-Weitzenb\"{o}ck flat $5D$ space-time on which we define a
5D vacuum in the Weitzenb\"{o}ck sense
\begin{equation}
^{(W)}R^a_{bcd}=0.
\end{equation}
Once this choice has been made, we can recover the Riemannian description by doing a transformation over the induced tensors (now obtained in terms of Weitzenb\"{o}ck connections)
according to $^{(LC)}\Gamma^{a}_{bc}=^{(W)}\Gamma^{a}_{bc}+K^{a}_{bc}$. Here $^{(LC)}\Gamma^{a}_{bc}$ is the symbol denoting the LC connection (the usual second king Christoffel symbols) and $K^{a}_{bc}$ is the contortion, which depends on the non-metricity and torsion.

In the context of WIMT we have studied gravito-electromagnetic
dynamics and magnetic monopoles on both, astrophysical and
cosmological scenarios\cite*{2}. The Weitzenb\"{o}ck torsion must
be viewed as a key element in order to construct a non-symmetric
connection capable to avoid the $d(d(A))=0$ dynamic condition, who
eliminates the magnetic monopoles from the theory. The cotangent
(penta-vector) $A$ is a 5D $1$-form characterized by
$\underrightarrow{A}=A_n\underrightarrow{e}^n$, $n=0...4$, with
components given by $A_0=\phi$ (the electric scalar potential),
$A_i=\mathbf{A}_i$ being the i-th component of the 3D vector
potential and $A_4\equiv\psi$ the gravitational scalar potential.
The gravito-magnetic dynamic equations in terms of exterior covariant derivatives
\begin{footnote}{A $p$-form is a tensor object which we call $W$ in present footnote, this is $p$ times cotangent and totally antisymmetric
\begin{eqnarray}\nonumber W=\frac{1}{p!}\,w_{i_1\,...\,i_p}\,\underrightarrow{e}^{i_1}\wedge...\wedge\underrightarrow{e}^{i_p},\end{eqnarray} in which wedge product is the complete anti-symmetrization of the tensorial product. The exterior covariant derivative is linked to a covariant derivative
by
\begin{eqnarray}\nonumber d(W)=\frac{1}{p!}\,w_{i_1\,...\,i_p\,;k}\,\underrightarrow{e}^k \wedge \underrightarrow{e}^{i_1}\wedge...\wedge\underrightarrow{e}^{i_p},\end{eqnarray} each different covariant derivative $;$ defines a different exterior derivative. The adjoint operation denoted by $*$ is defined in a manifold of dimension $m$
with the expression
\begin{eqnarray}\nonumber *W=\frac{\sqrt{|g|}}{(m-p)!\,p!}\varepsilon_{j_1\,...\,j_pi_{p+1}\,...\,i_n}w^{j_1\,...\,j_p}\,&\underbrace{
\underrightarrow{e}^{i_{p+1}}\wedge...\wedge\underrightarrow{e}^{i_m}}&,\\\nonumber &m-p&\end{eqnarray} then adjoint operation interchanges tensor order from $p$ to $m-p$. Taking last three equations we must write the magnetic current of the Maxwell equations as in \ref{maxwell}. For more details, the reader can see \cite*{gron}. }\end{footnote}, are
\begin{eqnarray}\label{maxwell}^{(m)}J=*d(F)=*d(d(A)),
\end{eqnarray}
with the Faraday $2$-form defined by $F=d(A)$. The asterisk
denotes the duality operator, which, in 5D, originates the apparition of
the $3$-form $*F$ which is the dual of $F$ in the expression
(\ref{maxwell}). This is very clear in next equations, especially
in Eq. (\ref{mono1}). On the other hand results obvious that
$d(d(A))=0$ implies $*d(d(A))=0$. Indeed, for a general affine
connection, we obtain
\begin{eqnarray}\label{1}
d(d(A))& := & d(A_{n;m}\underrightarrow{e}^m \wedge
\underrightarrow{e}^n)=\\\nonumber
&=&A_{n;m;p}\underrightarrow{e}^p \wedge \underrightarrow{e}^m
\wedge \underrightarrow{e}^n=\\\nonumber
&=&\{\overrightarrow{e}_p(\overrightarrow{e}_n(A_m))-\overrightarrow{e}_p(A_l\Gamma^l_{nm})\\\nonumber
&\,&-(\overrightarrow{e}_k(A_m)-A_l\Gamma^l_{km})\Gamma^k_{np}\nonumber \\
&-& (\overrightarrow{e}_n(A_k)-A_l\Gamma^l_{nk})\Gamma^k_{mp}\}\underrightarrow{e}^p
\wedge \underrightarrow{e}^m \wedge \underrightarrow{e}^n, \nonumber \\
\\
\label{mono1} *d(d(A)) & := & \left(\left[*F\right]_{abp}
\right)^{;p}\, e^a \wedge e^b \nonumber \\
& = & \left( \frac{\sqrt{|g|}}{2! 3!}
\epsilon_{abnmp} g^{ms} g^{pd} A^n_{; sd} \right) \, e^a \wedge
e^b,
\end{eqnarray}
where
\begin{eqnarray}
A^n_{\,;sd} &=& e_d \left(e_s (A^n)\right) + e_d \left( A^l \Gamma^n_{\,ls}\right) \nonumber \\
& + & \left( e_s \left(A^m\right)  +  A^l \Gamma^m_{ls}\right) \Gamma^n_{md} \nonumber \\
& - & \left( e_m (A^n) + A^l \Gamma^n_{lm} \right) \Gamma^m_{sd}.
\end{eqnarray}
It is easy to see that for a coordinate basis
$\{\overrightarrow{e}_n\}$, of the $5D$ tangent space $5DTM$, the
first term of Eq. (\ref{1}) vanishes, and the last three terms are
proportional to the product of the shape $\Gamma^l_{nm}
\,\underrightarrow{e}^n \wedge \underrightarrow{e}^m$, in which
the wedge product is $(n,m)$-antisymmetric. The entire term
results to be zero if the connection $\Gamma^l_{nm}$ is
$(n,m)$-symmetric (as is the case for the Christoffel symbols).
Then Eq. (\ref{1}) and Eq. (\ref{mono1}) are both zero, and in
such case there are no magnetic monopoles. In order to relax the
last condition has been proposed \cite*{diracbaez} that $F$ to be
not an exact $2$-form in the entire space-time, in order to obtain $F
\neq d(A)$ in some points where are located the magnetic
monopoles. In our case, for an 5D orthogonal coordinate basis, we keep the relationship
$^{(LC)} \nabla_b F^{nb}=J^n$, and the dual equations are $^{(LC)} \nabla_b
\left(*F\right)^{nmb}=0$. However, when we rewrite these equations in terms of a Weitzenb\"{o}ck geometric
representation, results so that these equations expressed in
terms of  divergence applied to the
Weitzenb\"{o}ck Maxwell tensor and its dual tensor inhomogeneous:
\begin{eqnarray}
^{(W)}\nabla_b ^{(W)} F^{nb} &= & ^{(ge)} J^{n}, \\
^{(W)}\nabla_c ^{(W)} {\cal F}^{abc} & = & ^{(gm)} J^{ab} ,
\end{eqnarray}
such that the gravito-electric and gravito-magnetic currents ($^{(ge)} J^{n}$, $^{(gm)} J^{ab}$), obtained from the Weitzenb\"{o}ck representation of a 5D gravito-electrodynamical theory, are given by
\begin{eqnarray}
^{(ge)} J^{n} &=& -\left[ \left(^{(W)} F^{mb} + g^{rm} A^p K^b_{\,\,pr} - g^{rb} A^p K^m_{\,\,pr} \right) K^n_{\,\,mb}  \right.\nonumber \\
& + & \left. \left( ^{(W)} F^{nm} +  g^{rn} A^p K^{m}_{\,\,pr} - g^{rm} A^p K^n_{\,\,pr} \right) K^b_{\,\,mb}\right], \nonumber \\
\\
^{(gm)} J^{ab} & = & \left[\frac{\sqrt{|g|}}{2!}\,\epsilon_{abcde}\,g^{ck}\,g^{dl}\,g^{es}\,\left\{(A_{n\,,l}-A_{l\,,n})\, K^n_{\,\,sk} \right. \right. \nonumber \\
&- &(A_{s\,,n}-A_{n\,,s})\, K^n_{\,\,lk}
 + \,A_m \left( K^m_{\,\,[sl]}+ K^m_{\,\,[nl]}\, ^{(W)} \Gamma^n_{\,\,sk} \right.\nonumber \\
 &+ & \left.\left.\left.  K^m_{\,\,[sn]}\, ^{(W)} \Gamma^n_{\,\,lk}\right)\right\}\right],
\end{eqnarray}
such that we can define the components of gravito-magnetic currents as $^{(gm)}
J^{a}= ^{(gm)} J^{ab}\, \bar{U}_b$, such that $\bar{U}_b$ are the
covariant components of the relativistic penta-velocity that
comply with: $\frac{d\bar{U}^a}{ds} + ^{(W)}\Gamma^a_{dc}
\,\bar{U}^{d} \bar{U}^c=0$. Therefore, the Maxwell tensor in the
Weitzenb\"{o}ck representation is: $^{(W)} F_{ab} = \,^{(W)}
\nabla_a A_b - \,^{(W)} \nabla_b A_a$, such that $^{(W)} \nabla_a
A_b = A_{b,a} - \, ^{(W)} \Gamma^c_{\,\,ab} A_c$, with $^{(W)}
\Gamma^c_{\,\,ab} = \,^{(LC)} \Gamma^c_{\,\,ab} - K^c_{\,\,ab}$, and
the dual Maxwell tensor in the Weitzenb\"{o}ck representation is
$^{(W)}{\cal F}^{nmb} = \frac{\sqrt{|g|}}{2!}\,\epsilon^{nmbac} \,
^{(W)}{F}_{ac}$. The contortion $K^c_{\,\,ab}$ is originated in
the fact that the Weitzenb\"{o}ck connections have an
antisymmetric contribution.

The Big Bang cosmology predicts that at very large number of heavy, stable magnetic monopoles should have been produced in the very early universe.
However, magnetic monopoles have never been observed, so if they exist at all, they are much rarer than the Big Bang theory predicts, or they should be in causally disconnected regions of the universe, as for example: inside of a black-hole. In this work we shall connect the Christoffel with the Weitzenb\"{o}ck geometrical
representations as was realised in a previous work\cite*{nuestro}, but extending this approach with the aim to introduce 4D magnetic monopoles in the Weitzenb\"{o}ck representation from a 5D Weitzenb\"{o}ck vacuum for a Dyonic Reissner Nordstr\"{o}m Black Hole (DRNBH). DRNBH differentiates from standard RNBH in that the first ones have both kind of charges, electric and magnetic, inside the horizon. As was demonstrated in \cite*{nuestro}, the imposed conditions over the 5D relativistic invariants there studied\begin{footnote}{
The 5D invariants proposed in \cite*{nuestro}, are
\begin{eqnarray}
 \label{inv00}^{(m)}\underrightarrow{\underrightarrow{J}}\left(\,^{(e)}\overrightarrow{J},\overrightarrow{U}\right)&:=&\,^{(e)}J^{a}\,^{(m)}J_{ab}\,U^{b}=\,^{(m)}J_{a} \,^{(e)}J^{a}, \label{1111} \\
 \label{inv01}^{(m)}J&:=&\,^{(m)}\underrightarrow{J}\left(\overrightarrow{U}\right)=\,^{(m)}J_{a}\,U^{a}, \label{2} \\
 \label{inv02}^{(e)}J&:=&\,^{(e)}\underrightarrow{J}\left(\overrightarrow{U}\right)=\,^{(e)}J_{a}\,U^{a}, \label{3} \\
 \label{inv03} ^{(gem)}J^2&:=&\left(^{ (e)}\underrightarrow{J} \, \wedge
\,^{(m)}\underrightarrow{J}\right)\left(^{(e)}\overrightarrow{J}\,\wedge\,^{(m)}\overrightarrow{J}\right)\nonumber \\
&=& \,^{(gem)}J_{ab}\,^{(gem)}J^{ab}, \label{4}
\end{eqnarray}
where $^{(gem)}\underrightarrow{\underrightarrow{J}}=\,^{(ge)}\underrightarrow{J} \,\wedge \,^{(gm)}\underrightarrow{J}$ are the gravito-electro-magnetic $2$-form, and $U^{b}$ are the components of the penta-velocities of the observers.}
\end{footnote},
prohibit the existence of a DRNBH, unless that densities of gravito-magnetic mass and the gravito-electric charges to be equals: $\rho_M = \rho_e$, so that they only admit naked singularities. For this reason, we shall try to solve this problem by treating the invariants proposed in \cite*{nuestro} as a particular set of values, which must be obtained from an invariant arising from the geometric product of the currents.

\section{Dyonic Reissner-Nordstr\"{o}m Black Hole (DRNBH) and extended Dirac quantization from geometric product.}

To address the problem of a DRNBH our propose is to start from an extended $5D$ DRNBH metric, given by
\begin{eqnarray}\label{drnbh5} ^{(5D)}ds^2&=&f(r)\,dt^2- \left[f(r)\right]^{-1}dr^2 \nonumber \\
&-& r^2 \left(d\theta^2- {\sin\theta}^2 d\varphi^2\right)-dl^2,
\end{eqnarray}
where $f(r)=\left[1-\frac{2M}{r}+\frac{({Q_e}^2+{Q_m}^2)}{r^2}\right]$. This metric does not provide us a 5D Riemann vacuum, but is a Weitzenb\"{o}c one.
The effective $4D$ DRNBH\cite*{rn} is obtained by making a constant foliation over the extra dimension: $l=l_0$
\begin{eqnarray}\nonumber
^{(4D)}ds^2&=&f(r)\,dt^2- \left[f(r)\right]^{-1}dr^2 \nonumber \\
&-& r^2 \left(d\theta^2- {\sin\theta}^2 d\varphi^2\right),
\end{eqnarray}
such that the radiuses are
\begin{eqnarray}\label{radiod0}
r_{\pm}=M\pm \sqrt{M^2-\left[{Q_e}^2+{Q_m}^2\right]},
\end{eqnarray}
such that the inner horizon is in $r_{-}$.
In order to define grovito-electromagnetic invariants with the sources (or currents), we shall think $^{(ge)}\underrightarrow{J}\,^{(gm)}\underrightarrow{J}$ as a geometric product given by both, inner and wedge products of $^{(ge)}\underrightarrow{J}$ and $^{(gm)}\underrightarrow{J}$. The first one is the symmetric part of the geometric product and the second one is its antisymmetric part\cite*{desabata}:
\begin{eqnarray}\label{product0}
^{(ge)}\underrightarrow{J}\,^{(gm)}\underrightarrow{J}=\,^{(ge)}\underrightarrow{J}\cdot\,^{(gm)}\underrightarrow{J}+\,^{(ge)}\underrightarrow{J}\wedge\,^{(gm)}\underrightarrow{J}.
\end{eqnarray}
The first term in the right side is a scalar and the second one is a $2$-form (or bi-covector). In general, we must write a scalar according to
\begin{footnote}{For a general multi-tensor object
\begin{eqnarray}\label{t}
T & = & A+B_n \underrightarrow{e}^n+C_{nm} \underrightarrow{e}^n \otimes  \underrightarrow{e}^m  \nonumber \\
& + & D_{nmp}\underrightarrow{e}^n \otimes  \underrightarrow{e}^m \otimes \underrightarrow{e}^p+...
\end{eqnarray}
where $A, B_n, C_{nm}, D_{nmp}$ are arbitrary scalar functions in $\mathfrak{F}(M)$. We must obtain a scalar defined by
\begin{eqnarray}\label{t2}
T^2&=&A^2+ B_n B_{n'} \overrightarrow{\overrightarrow{g}}(\underrightarrow{e}^n,\underrightarrow{e}^{n'})\nonumber \\
&+& C_{nm} C_{n'm'} \overrightarrow{\overrightarrow{g}}(\underrightarrow{e}^n,\underrightarrow{e}^{n'}) \overrightarrow{\overrightarrow{g}}(\underrightarrow{e}^m,\underrightarrow{e}^{m'})  \nonumber \\
&+&  D_{nmp} D_{n'm'p'} \overrightarrow{\overrightarrow{g}}(\underrightarrow{e}^n,\underrightarrow{e}^{n'}) \overrightarrow{\overrightarrow{g}}(\underrightarrow{e}^m,\underrightarrow{e}^{m'}) \overrightarrow{\overrightarrow{g}}(\underrightarrow{e}^p,\underrightarrow{e}^{p'}) + ...\nonumber \\
&=& A^2 + B_n B_{n'} g^{nn'}+ C_{nm} C_{n'm'} g^{nn'} g^{mm'} \nonumber \\
&+ & D_{nmp} D_{n'm'p'} g^{nn'} g^{mm'} g^{pp'}+... ,
\end{eqnarray}
which is a generalized expression of a inner product for multi-tensorial objects of any kind.
We must notice that such product coincides with the usual inner product for a pure vector or co-vector.}
\end{footnote}
\begin{eqnarray}\label{product1}
(^{(ge)}\underrightarrow{J}\,^{(gm)}\underrightarrow{J})^2&=&\overrightarrow{\overrightarrow{g}}(\,^{(ge)}\underrightarrow{J},\,^{(gm)}\underrightarrow{J})^2
+(^{(ge)}\underrightarrow{J}\wedge\,^{(gm)}\underrightarrow{J})^2 \nonumber \\
&=&(^{(ge)}J_n\,^{(gm)}J^n)^2+\,^{(gem)}J_{nm}\,^{(gem)}J^{nm}, \nonumber \\
& &
\end{eqnarray}
which is a particular application of Eq. (\ref{t2}). In \cite*{nuestro} was proposed the that
\begin{eqnarray}\label{product2}
\underbrace{(^{(ge)}\underrightarrow{J}\,^{(gm)}\underrightarrow{J})^2}&=&\underbrace{(^{(ge)}J_n\,^{(gm)}J^n)^2}+\underbrace{\,^{(gem)}J_{nm}\,^{(gem)}J^{nm}},\nonumber \\ \,\,\,\,\,\,\,\,\,\,\,\,\,\,\,\,\,\,n^2\,\,\,\,\,\,\,\,\,\,\,\,\,&\,&\,\,\,\,\,\,\,\,\,\,\,\,\,\,\,0\,\,\,\,\,\,\,\,\,\,\,\,\,\,\,\,\,\,\,\,\,\,\,\,\,\,\,\,\,\,\,\,\,\,\,\,\,\,\,\,\,\,\,\,n^2. \nonumber \\
& &
\end{eqnarray}
This proposal works good when the gravito-electric and gravito-magnetic charges are in space-time regions causally disconnected, as is the case of a RNBH. However, there are problems in cases when the charges co-exist in causally connected regions of the space-time. This is the case in a DRNBH when $Q_m\neq 0$. This suggests that,
in cases when gravito-magnetic and gravito-electric charges co-exist in the same causally connected space-time, we
must require that the co-vector currents must not be orthogonal
\begin{eqnarray}\label{m2}
^{(ge)}J_n\,^{(gm)}J^n\neq 0.
\end{eqnarray}
Therefore, in the cases where the source has both, gravito-magnetic and gravito-electric charges, the currents cannot be considered as orthogonal, and
\begin{eqnarray}\label{product3}
(^{(ge)}\underrightarrow{J}\,^{(gm)}\underrightarrow{J})^2=n^2,
\end{eqnarray}
According to Eq. (\ref{m2}), the first term in r.h.s. of (\ref{product2}) is the square of a simple scalar let's say $(^{(ge)}J_n\,^{(gm)}J^n)^2=m^2$, and using this expression
with (\ref{product3}), in (\ref{product2})\footnote{The reader can see the Sect. {\bf 3} of \cite*{nuestro}, where we demonstrated that
\begin{eqnarray}\nonumber
[^{(gm)}J_{A}]=(-\rho_m \, U^5,0,0,0,\rho_m \, U^1), \qquad {\rm with} \qquad ^{(gm)}J_i=0,
\end{eqnarray}
and
\begin{eqnarray}\label{je5}
[^{(ge)}J_{A}]=(\rho_e,0,0,0,\rho_M).
\end{eqnarray}},
we obtain
\begin{eqnarray}\label{dirac1}
\rho_e\rho_m=\frac{\sqrt{n^2-m^2}}{2},
\end{eqnarray}
in which $n^2$ is the square of the norm of the geometric product of the currents and $m^2$ is nonzero due to the non-orthogonality of such currents.
If the gravito-electric and gravito-magnetic currents were mutually orthogonal then would be $m^2=0$ and (\ref{dirac1}) would reduce to the already known quantization relation obtained in\cite*{nuestro}. In general, if we assume that $U^1,\,U^5 \neq 0$, we obtain
\begin{eqnarray}\label{roe}
\rho_e&=&\frac{m^2 U^5}{{(U^5)}^2-{(U^1)}^2},\\\label{roM}\rho_M&=&\frac{m^2 U^1}{{(U^5)}^2-{(U^1)}^2},\\\label{rom}\rho_m&=&\frac{\sqrt{n^2-m^2}}{2}\left[\frac{{(U^5)}^2-{(U^1)}^2}{m^2 U^5}\right].
\end{eqnarray}
Since $M^2-Q_e^2>0$, one must require $|U^1|>|U^5|$, so that this condition enables us to obtain $\rho_m\neq 0$ in the presence of a true DRNBH
\begin{footnote}{The
radius remains real for $\frac{4m^8}{n^2-m^2}>\frac{{U^5}^2-{U^1}^2}{U^5}$, which reduces to $\frac{4m^8}{n^2-m^2}>1$ over the horizon.}
\end{footnote}.
Hence, for a 4D space-time obtained by a constant foliation
\begin{footnote}{The Latin index $n,m,p=1...5$ are associated to the 5D manifold taking values along the five coordinates, the Greek index $\alpha,\beta=1...4$ are associated to the effective 4D space-time taking values along the four effective coordinates obtained after the foliation.}
\end{footnote},
it holds that $g_{\alpha\beta}U^\alpha U^\beta=g_{11}U^1U^1=1$. Then, the induced four-velocity is completely given by $U^1=\sqrt{1-\frac{2M}{r}+\frac{(Q_e^2+Q_m^2)}{r^2}}$. On the other hand, from the corresponding 5D geodesic equation for the 5D DRNBH-like metric, we obtain that $U^5=K(s)$, which is a constant of the affine parameter.

\section{Final comments}

We have extended the charge quantization prescription suggested in \cite*{nuestro}, by setting a geometric invariant given by the square of a 5D geometric product defined in (\ref{product1}).
We have obtained that, under an extension of (\ref{product2}), given by the quantization condition (\ref{dirac1}), provides a relationship between the gravito-electric, the gravito-magnetic and the gravitational mass densities,
which can take discrete values according to (\ref{roe}), (\ref{roM}) and (\ref{rom}). They lead us to a set of conditions compatible with an effective 4D DRNBH.
We must remark that in our previous work\cite*{nuestro} we check that the obtained gravito-magnetic currents are in presence of a stable gravito-magnetic monopole which is inside the RNBH, but the relativistic observer which is placed outside the BH is not sensitive to the gravito-magnetic charge of the BH, because they (the gravito-electric and gravito-magnetic charges) are causally disconnected. For this reason in that case the gravito-electric and gravito-magnetic currents are orthogonal. However, in this work we have studied a different reality by dealing with a DRNBH in which both, $Q_e$ and $Q_m$, coexist in a same causally connected region of space-time and for this reason $^{(ge)}\underrightarrow{J}$ and $^{(gm)} \overrightarrow{J}$ cannot be considered as orthogonal, as is the case of a RNBH. Of course, the study of duality between the interior and exterior space-time, and their physical properties, deserves a more profound study.

\end{document}